\documentclass[a4paper]{jpconf}
\usepackage{graphicx}
\newcommand{\gep}{G_{Ep}}
\newcommand{\gmp}{G_{Mp}}
\newcommand{\gmn}{G_{Mn}}
\newcommand{\gmpmu}{G_{Mp}/\mu_{p}}

\newcommand{\gen}{G_{En}}
\newcommand{\gmnmu}{G_{Mn}/\mu_{n}}

\newcommand{\numu}{\nu_{\mu}}
\newcommand{\muminus}{\mu^{-}}

\newcommand{\numubar}{\overline{\nu}_{\mu}}

\begin{document}

\title{Extraction of the  Axial
Nucleon Form Factor from Neutrino Experiments on Deuterium}
\author{ A. Bodek, S. Avvakumov, R. Bradford, and  H. Budd}
\address{Department of Physics and Astronomy, University of Rochester, Rochester, NY  14627-0171}


\begin{abstract}
We present new
parameterizations of vector and axial nucleon form factors.
We maintain an excellent descriptions of the
form factors at low momentum transfers ($Q^2$), where the spatial
structure of the nucleon is important, and
use the Nachtman scaling variable $\xi$
to relate elastic and inelastic form factors
and impose quark-hadron duality constraints at high $Q^2$
 where the quark structure dominates. 
We use the new vector form factors to re-extract
updated values of the axial form factor from  $\numu$
experiments on deuterium.
 We obtain an updated world
average value from $\numu$d, $\numubar$H
and pion electroproduction experiments of 
$M_{A}$ = $1.014 \pm 0.014~GeV/c^2$. Our parameterizations are useful 
in modeling $\nu$ interactions at low energies
(e.g. for $\numu$ oscillations experiments).  The predictions
for high $Q^2$  can be tested in the 
next generation electron and $\numu$ scattering experiments. (Presented by A. Bodek at the European Physical Society Meeting,  EPS2007, Manchester, England, July 2007).
\end{abstract}

The nucleon vector and axial elastic form factors have been measured
for more than 50 years in $e^- N$ and $\nu N$ scattering.
At low $Q^2$, a reasonable  description of the proton and neutron
elastic form factors is given by the dipole approximation.
The dipole approximation is a lowest-order attempt to
incorporate the non-zero size of the proton into the form
factors. The approximation assumes that the proton has
a simple exponential spatial charge distribution, $\rho(r)=\rho_0
e^{-r/r_0}$, where $r_0$ is the scale of the proton radius.
Since the form factors are related in
the non-relativistic limit to the Fourier transform of the
charge and magnetic moment distribution, the above $\rho(r)$
yields the dipole form defined by:  $G_D^{V,A}(Q^2) =
{C^{V,A}}/{\left(1+\frac{Q^2}{M_{V,A}^2}\right)^2} $.
Here $C^{V,A}$= (1,$g_{A}$), $g_{A}$ = -1.267, $M_{V}^2$ = 0.71 $(GeV/c)^2$,
  and $M_{A}$ = 1.015 $GeV/c^{2}$ (see below). 
  
Since $M_{A}$ is not equal to $M_{V}$, the 
distribution of electric and axial charge are different.
However, the magnetic moment distributions were assumed to have the
same spatial dependence as the charge distribution ({\it
i.e.}, form factor scaling). Recent measurements from Jefferson Lab 
show that the ratio of $\frac{\mu _p G_{Ep}}{G_{Mp}}$ 
falls at high $Q^2$
challenging the validity of form factor scaling
and  resulting in new updated parameterizations of
the form factors
\cite{kelly}). 
In this paper we present
parameterizations that simultaneously satisfy constraints at low $Q^2$ where the
spatial structure of the nucleon is important, and at high
$Q^2$ where the quark structure is important. A violation of
form-factor scaling is expected from quark-hadron
duality.
We use our
new vector form factors to re-extract updated values of the axial form factor
from a re-analysis of previous neutrino scattering data 
on deuterium and present a new parameterization for the axial form
factor within the framework of quark-hadron duality.

The new parameterizations presented in this paper are referred to as the duality
based ``BBBA07''
parameterization. Our updated parameterizations feature the following:
(1) Improved functional form that adds an additional 
$Q^2$ dependence using the Nachtman scaling variable $\xi$
to relate elastic and inelastic
form factors.
For elastic scattering ($x=1$) 
$\xi^{p,n,N}=\frac{2}{(1+\sqrt{1+1/\tau_{p,n,N}})}$, where $\tau_{p,n,N} =
Q^2/4M_{p,n,N}^2$. Here
$M_{p,n,N}$ are the proton (0.9383 $GeV/c^2$), neutron (0.9396 $GeV/c^2$), 
and average nucleon mass (for proton,
neutron, and axial form factors, respectively).
(2) Yield the same values as
Arrington and Sick \cite{arringtonsick} for $Q^{2}< 0.64
(GeV/c)^{2}$,
while satisfying 
quark-hadron duality constraints at high-$Q^2$.

For vector form factors our fit functions are $A_N(\xi)$ (i.e. 
$A_{Ep}(\xi^{p})$, $A_{Mp}(\xi^{p})$, $A_{En}(\xi^{n})$,
$A_{Mn}(\xi^{n})$)
multiplying an updated Kelly\cite{kelly} type parameterization
 of one of the proton form factors. 
The Kelly parameterization is: $ G^{Kelly}(Q^2) = {\sum_{k=0}^{m}a_k \tau^k_{p}}/{1 +
\sum_{k=1}^{m+2}b_k \tau^k_{p}}$, 
%
where $a_{0}=1$ and $m=1$.
In our analysis, we use all the  
datasets used by Kelly\cite{kelly}, updated to include the
recent BLAST results, to
fit $\gep$, $\gen$, $\gmpmu$, and $\gmnmu$ ($\mu_p = 2.7928$, $\mu_n = -
1.9130$).
Our parameterization employs the published Kelly
functional form to $G^{Kelly}_{Ep}$, and an updated set of parameters
for $G^{Kelly-upd}_{MP}(Q^2)$.
The  parameters   $A_N(\xi)$ is given by 
\begin{eqnarray}
 A_N (\xi) &=& \sum_{j=1}^{n}   p_j\prod_{k=1, k \ne j}^{n} \frac{\xi - \xi_k}{\xi_j -
\xi_k}. \nonumber
\end{eqnarray}
  The $\xi_j$ are equidistant ``nodes'' on an interval $[0,1]$ and
$p_j$ are the
fit parameters that have an additional property $A_N (\xi_j) = p_j$.
The functional form $A_N(\xi)$ (for  $\gep$, $\gmp$, $\gen$, and $\gmn$)
is used with seven $p_j$ parameters
at $\xi_j$=0, 1/6, 1/3, 1/2, 2/3, 5/6, and 1.0.
In the fitting procedure  the parameters of $A_N(\xi)$ are
constrained 
to give the same vector form factors as the recent low $Q^2$
fit of Arrington and Sick \cite{arringtonsick} for $Q^{2}< 0.64
(GeV/c)^{2}$ (as that analysis includes coulombs corrections which 
modify $G_{Ep}$, and two photon exchange corrections which modify
$G_{Mp}$ and $G_{Mn}$). Our fits to the  form factors are:
\begin{eqnarray}
      {G_{Mp}(Q^2)}/{ \mu_{p}} &=& { A_{Mp}(\xi^{p})} \times 
      {G^{Kelly-upd}_{Mp}(Q^2)} \nonumber \\
 {G_{Ep}(Q^2)} &=& A_{Ep}(\xi^{p})\times {G^{Kelly}_{Ep}(Q^2)}
 \nonumber \\
    {G_{Mn}(Q^2)}/{\mu_{n}} &=& A^{25,43}_{Mn}(\xi^{n})
    \times  {G_{Mp}(Q^2)}/ {\mu_{p}}
 \nonumber \\   
   {G_{En}(Q^2)} &=& A^{25,43}_{En}(\xi^{n})\times {G_{Ep}(Q^2)} \times
   \left( {\frac{a\tau_{n}}{1+b\tau_{n}}} \right)
  \nonumber,
  \end{eqnarray}
 where we use 
 our updated parameters in the Kelly parameterizations.
 For $\gen$ the parameters a=1.7 and b=3.3 are the same as
 in the Galster\cite{kelly}  parametrization and ensure that
 $d\gen/dQ^2$ at for $Q^{2}=0$ is in agreement with measurements.
  The values  $A(\xi)$=$p_1$ at $\xi_1$=0  ($Q^2 =0$) for
   $\gmp$,  $\gep$,   $\gen$,   $\gmn$ are set to  
  to 1.0. The value  $A(\xi)$=$p_7$ at $\xi_j$=1 
  ($Q^2 \rightarrow \infty$) for $\gmp$ and $\gep$ is set to 1.0.
   The value  $A(\xi)$=$p_j$ at $\xi_j$=1 
for   $\gmn$ and  $\gen$ are fixed by  constraints from quark-hadron
duality\cite{bbba2007}.The parameters and plots of the  new form factors $\gep$, $\gmpmu$, $\gmnmu$, and $\gen$ are given in ref.\cite{bbba2007}.

\begin{table} 
\begin{center}
\begin{tabular}{|l|l|l|l|l|l|l|l| |
    }
    \hline 
$Experiment$ & QE & $Q^{2}$ range  &    $M_A$& $\Delta M_A$  &$M_A^{updated}$	\\
$\numu$d $\rightarrow$$\muminus$ p $p_{s}$ & events & $GeV/c^2$ &  
(published) &  FF &  $GeV/c^2$   \\
\hline 
$Mann_{73}$  & $166$ &  $.05-1.6$ &  0.95 $\pm$ .12   
&   &  \\
$Barish_{77}$   & $500$ &  $.05-1.6$ &  0.95 $\pm$ .09       & $-.026$ &   \\
    $Miller_{82,77,73}$ & $1737$ &  $.05-2.5$ &  1.00 $\pm$ .05      & $-.030$ &
    0.970 $\pm$ .05 \\
    \hline
     $Baker_{81}$ & $1138$ &  $.06-3.0$ &  1.07 $\pm$ .06 
     & $-.028$ &
     1.042 $\pm$ .06\\
     \hline 
 $Kitagaki_{83}$ & $362$ &  $.11-3.0$ &    1.05$_{-.16}^{+.12}$ & $-.025$
 & 1.025$_{-.16}^{+.12}$   \\ 
 \hline
 $Kitagaki_{90}$ & $2544$ &  $.10-3.0$ &  1.070$_{-.045}^{+.040}$ & $-.036$
 & 1.034$_{-.045}^{+.040}$\\ 
 \hline 
  $Allasia_{90}$    & 552 & .1-3.75 &  
 $1.080 \pm .08$ 
 &$-.080$ & $1.00 \pm .08$ \\
 \hline \hline
 Av.  $\numu$d      & 5780 & above & 
 & &  $1.014 \pm .026$ \\  \hline
   $\pi$~$electrprod.$      &  &   &    &  &  $1.014 \pm .016$ \\
  \hline \hline
    $\numubar$H $\rightarrow$$\muminus$ n     & 13 & 0-1.0 &
 $0.9\pm 0.35$ &$-.070$&  $0.83 \pm .35$ \\
 $\numubar$H $\rightarrow$$\muminus$ n     & 13 & 0-1.0 &
   $\sigma_{QE}$
  & &  $1.04 \pm .40$ \\
\hline
   $Average- all$      &  &   &    &  &  $1.014 \pm .014$ \\
 \hline
\end{tabular}
\end{center}
\caption{$M_A$ $(GeV/c^2)$ values published
by  $\numu$-deuterium experiments\cite{neutrinoD2} 
and updated corrections  $\Delta M_A$ when re-extracted with updated
$BBBA2007_{25}$\cite{bbba2007} form factors, and $g_{a}$=-1.267. Also 
shown is updated $M_{A}$ from $\numubar$Hydrogen 
$\rightarrow$$\muminus$ n \cite{hydrogen}. 
}
\label{MA_values}
\end{table}
 \begin{figure}
 \begin{center}
\includegraphics[width=6.0in,height=2.4in]{{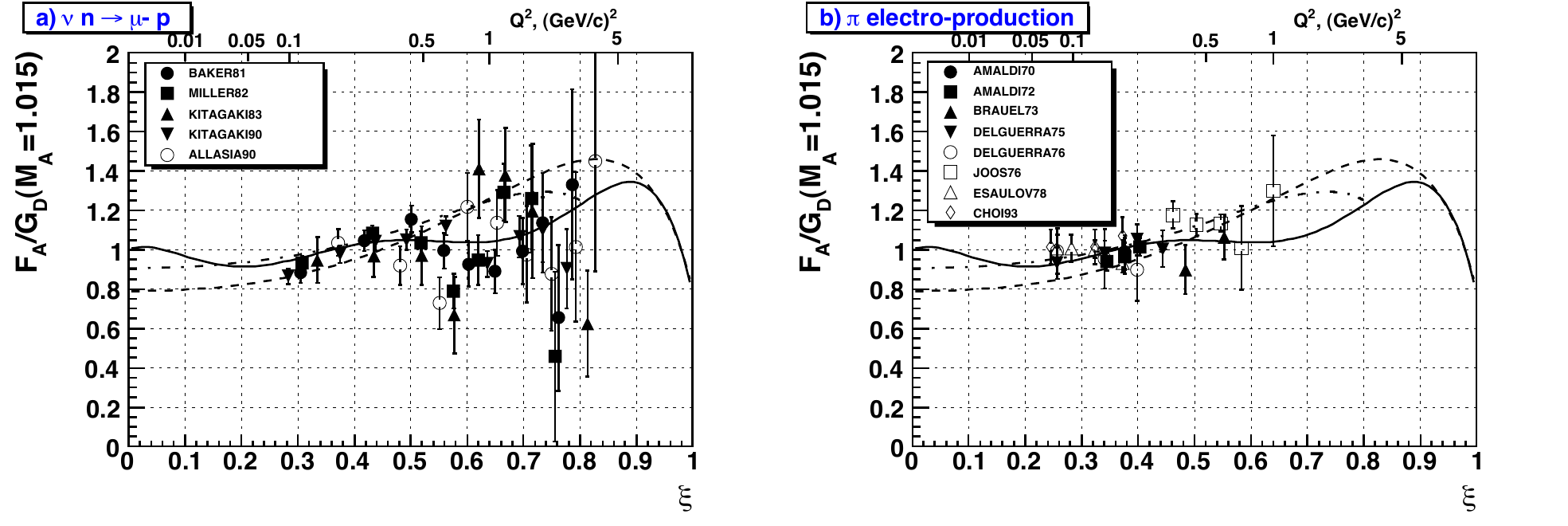}}
\caption[$F_{A}$:Axial Form factor ratios to $G_D^{A}$]{
(a) $F_A(Q^2)$ re-extracted from neutrino-deuterium
 data divided by $G_D^{A}(Q^2).$
 (b)   $F_A(Q^2)$ from pion electroproduction
divided by $G_D^{A}(Q^2)$,
 corrected for  for hadronic effects.
   Solid line - duality based  fit; Short-dashed line - $F_A(Q^2)_{A2=V2}$. 
The long-dashed line is $F_A(Q^2)_{A1=V1}$.
 Dashed-dot line - constituent quark model
 }
\end{center}
\end{figure}

Using our updated $BBBA2007_{25}$ form
factors and an updated
value $g_{A}$ = -1.267, we perform a complete
reanalysis of  published $\nu$ quasielastic \cite{neutrinoD2} (QE) data on
deuterium ($\numu$ n $\rightarrow$ $\muminus$p) using the procedure described in detail
in ref. \cite{deuterium}. 
We extract new values of $M_A$ (given in
Table \ref{MA_values}), and updated values of $F_A(Q^2)$ . The
average of the corrected measurements of $M_{A}$ from Table \ref{MA_values}
is  $1.0137 \pm 0.0264$  $GeV/c^2$.  This is to be compared to the average value
of $1.0140 \pm 0.0160$  $GeV/c^2$ extracted from pion electroproduction
experiments  after corrections for hadronic effects.  world
average value from $\numu$d, $\numubar$H
and pion electroproduction experiments of 
$M_{A}$ = $1.0137 \pm 0.0137~GeV/c^2$.  This is smaller than
the recent results\cite{boone} from MiniBoone on a carbon target 
($M_{A} = 1.23 \pm 0.20$ $GeV/c^2$) and
$K2K$\cite{k2k} on oxygen ($M_{A} = 1.20 \pm 0.12$
 $GeV/c^2$). Both collaborations use
updated vector form factors. collaborations attribute a difference from 
deuterium to nuclear effects. However, there is experimental
and theoretical evidence\cite{ma-nuclear} that 
 $M_{A}$ in nuclear targets is the same (or smaller)
 than in deuterium. 
 This 
 discrepancy is 
important for $\nu$  oscillations experiments since it
affects the normalization 
(at high energies the QE cross section
is approximately proportional to $M_{A}$) and non-linearity of the
QE cross section, which is relevant
to the extraction of  $\nu$ mass difference and mixing angle.

  For deep-inelastic scattering, the vector and axial
  parts of $F_{2}$ are equal. 
   Local quark-hadron duality 
    at large $Q^2$ implies  that
  the axial and vector parts of $F_{2}^{elastic}$
  are also equal: ${[F_A(Q^2)_{A2=V2}]^{2}} ={(G_E^{V})^{2}(Q^2)+\tau_{N} 
(G_M^V(Q^2))^{2}}/{(1 + \tau_{N})},$
%
where $ G_E^V(Q^2)=G_{Ep}(Q^2)-G_{En}(Q^2) $ and  
  $G_M^{V}(Q^2) = G_{Mp}(Q^2)-G_{Mn}(Q^2)$.
 
We extract values of $F_A(Q^2)$  from
the differential
 cross sections using the procedure
 of ref. \cite{deuterium}.
 The overall normalization is set by the theoretical QE
 cross section\cite{GD}.
  We  then do a duality based fit
  to 
 $F_A(Q^2)$ (including pion
 electroproduction data) 
 of the form: $F_A (Q^2)=A^{25}_{FA} (\xi^{N}) \times G_D^{A}(Q^2).$

\begin{figure}
 \begin{center}
\includegraphics[width=6.0in,height=2.4in]{{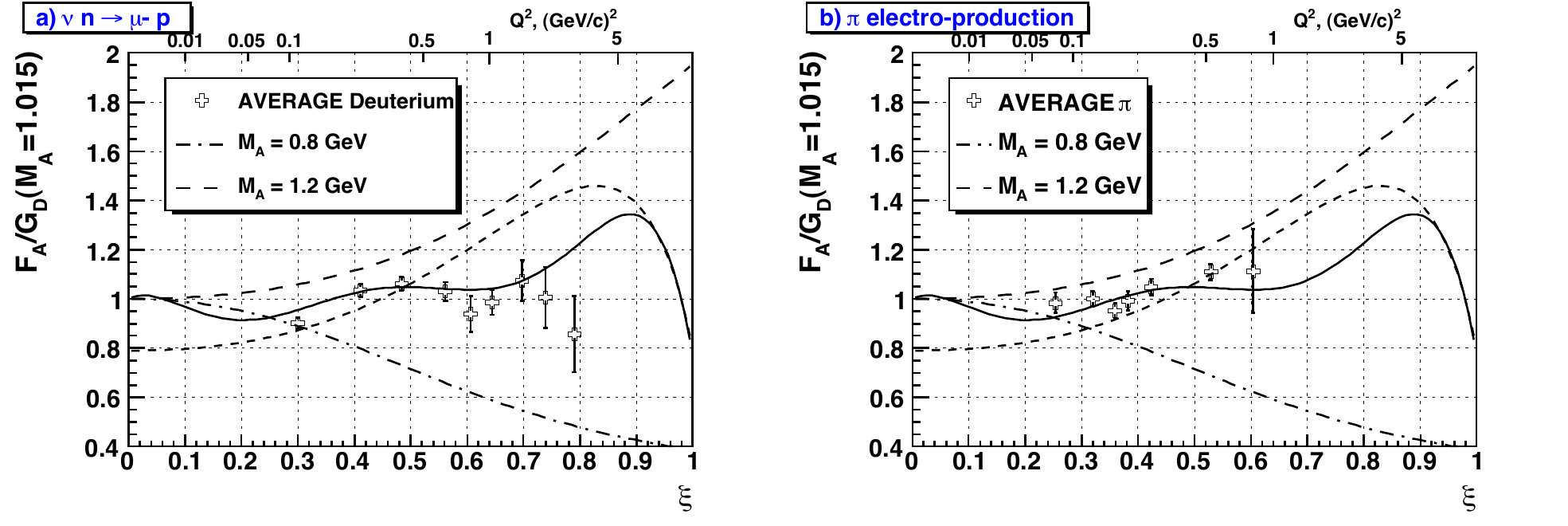}}
\caption[$F_{A}$:Average ratios to $G_D^{A}$]{
Average $F_A(Q^2)$/$G_{D}(M_{A}=1.015)$  from (a) $\nu$-deuterium 
 experiments, and  (b) from pion electroproduction experiments.
 Also shown are  the average E734 $\numu$ and $\numubar$ results from carbon. 
 Solid line - duality based
  fit; Short-dashed line - $F_A(Q^2)_{A2=V2}$; 
   long-dashed)- $G_{D}(M_{A}=1.20)/G_{D}(M_{A}=1.015)$; dashed-dot
  line - $G_{D}(M_{A}=0.80)/G_{D}(M_{A}=1.015)$
 }
\end{center}
\end{figure}

We impose the constraint  $A^{25}_{FA} (\xi_1=0) = p_1 = 1.0$. We also
constrain the fit
by requiring
that $A^{25}_{FA}(\xi^{N})$ yield  
$F_A(Q^2)=F_A(Q^2)_{A2=V2}$ by including additional $''fake''$ data
points) for $\xi > 0.9$  ($Q^{2} >7.2 (GeV/c)^2$).
 Figure 1(a) shows 
  $F_A(Q^2)$  extracted from neutrino-deuterium experiments 
  divided by $G_D^{A}(Q^2)$\cite{GD}.
Figure 1(b)  shows 
 $F_A(Q^2)$  extracted from pion electroproduction
 experiments 
  divided by $G_D^{A}(Q^2)$\cite{GD}. These pion electroproduction
  values can be directly compared to the neutrino results because they
  are  multiplied  by a factor
  $F_{A}(Q^{2},M_{A}=1.014~GeV/c^{2})$/$F_{A}(Q^{2},M_{A}=1.069~GeV/c^{2})$  
  to correct  for $\Delta M_{A} = 0.055~GeV/c^{2}$ originating from 
   hadronic effects. Figure (2) shows the same data averaged over
   all experiments in bins of  $Q^2$.
    The solid line is our duality based
 fit. The short-dashed line is  $F_A(Q^2)_{A2=V2}$. 
  The dashed-dot line
  is a constituent-quark model\cite{quark} prediction.
 
In summary, our new parameterizations
are useful in modeling $\nu$ interactions
for oscillations experiments.
Our predictions
for $\gen (Q^{2}) $ and $F_{A}(Q^{2})$ 
can be tested in future $e-N$ and $\nu$-N
 experiments.


\begin{thebibliography}{20}


   \bibitem{bbba2007} A. Bodek, S. Avvakumov, R. bradford, and H.
    Budd, hep-ex/0708.1946;  $www.pas.rochester.edu/\sim~bodek/FF/$



\bibitem{kelly}J.J. Kelly, Phys. Rev. C 70, 068202 (2004);  S. Galster \textit{et al}, Nucl. Phys. B32, 221
(1971).




\bibitem{arringtonsick} J. Arrington and  I.Sick nucl-th/0612079






\bibitem{deuterium} H. Budd, A. Bodek, J. Arrington,
Nucl.Phys.Proc.Suppl. 139, 90 (2005).

\bibitem{pion}V. Bernard \textit{et al}, J. Phys. G28, R1 (2002). 

\bibitem{neutrinoD2}
W.A. Mann et al Phys. Rev. Lett. 16, 3103 (1973);
S.J.~Barish {\em et al.}, Phys. Rev. D16 (1977) 3103;
K.L.~Miller {\em et al.}, Phys. Rev. D26 (1982) 537;
N.J.~Baker {\em et al.}, Phys. Rev. D23 (1981)
2499; 
T.~Kitagaki {\em et al.}, Phys. Rev. D28 (1983)
436; T.~Kitagaki {\em et al.}, Phys. Rev. D42 (1990)  1331;
 D. Allasia {\em et al.}, Nucl. Phys. B343 (1990) 285.
\bibitem{hydrogen}G. Fanourakis {\em et al.},Phys. Rev. D21 (1980) 562.

\bibitem{quark} R. F., Wegenbrunn, \textit{et al}, hep-ph/0212190.

\bibitem{boone} A. Aguilar-Areval \textit{et al} (MiniBoone)
hep-ex/0706.0926

\bibitem{k2k} R. Gran \textit {et al} (K2K), Phys. Rev. D74 (2006)
052002.

\bibitem{ma-nuclear} S. K. Singh and  E. Oset, Nucl.Phys A542 (1992) 587. 



\bibitem{GD} Evaluated with $M_{A} = 1.015~GeV/c^{2}$.


\end{thebibliography}
\end{document}